\documentclass{osa-article}

\journal{osajournal}
\usepackage{nicefrac}
\articletype{Research Article}
\newenvironment{myindentpar}[1]%
  {\begin{list}{}%
          {\setlength{\leftmargin}{#1}}%
          \item[]%
  }
  {\end{list}}
\begin{document}

\title{Efficient side-coupling to photonic crystal nanobeam cavities via state-space overlap}

\author{Francis O. Afzal,\authormark{*} Sami I. Halimi,\authormark{} and Sharon M. Weiss\authormark{}}

\address{\authormark{} Department of Electrical Engineering, Vanderbilt University, 2201 West End Ave, Nashville, TN 37235, USA\\}

\email{\authormark{*}francis.afzal@vanderbilt.edu} %% email address is required



\begin{abstract}
We present design guidelines for realizing side-coupled photonic crystal nanobeam (PCN) cavities with efficient coupling to low order resonances through adjusting the overlap of the PCN cavity mode and a feeding bus waveguide in both physical and k-space. We show that optimal side-coupled configurations function at a non-zero k-vector offset between the bus waveguide and PCN cavity modes. The straightforward design of side-coupled PCNs with high contrast resonances opens the door to their practical implementation in multiplexed, on-chip photonic devices.
\end{abstract}

%%%%%%%%%%%%%%%%%%%%%%%%%%  body  %%%%%%%%%%%%%%%%%%%%%%%%%%
\section{Introduction}
Optically resonant nanophotonic structures hold great promise in applications including optical signal processing \cite{LNmodulator,grapheneringres,squeezedlight}, nanomanipulation \cite{plasmonictrapping,EricksonNanomanipulation,ShapeableOpticalTweezers}, photovoltaics \cite{solarcell}, imaging \cite{Metalens}, and sensing \cite{EricksonSensorArrays,SingleNanoparticleDetection}. On-resonance, these structures support high, localized optical field intensities that can be leveraged for enhancing various types of interactions between light and matter. Resonant field enhancements in photonic and plasmonic devices result directly from the confinement of light both spatially, as characterized by the mode volume ($V$), and temporally, as characterized by the quality factor ($Q$). Most resonant nanophotonic structures have the ability to attain either high $Q$ or low $V$, but lack the ability to simultaneously demonstrate significant spatial and temporal confinement (i.e., high $Q/V$) \cite{plasmonictrapping,PlasmonicNanodisk, HighQWhisperingGallery,WhisperingGalleryLowModeVolume}. Photonic crystal nanobeam (PCN) cavities are the exception \cite{JudsonSlot,UltraSlot,OpMechHighQ/V,ShurenArxiv}. PCN cavities have the design flexibility necessary to achieve low $V$ while maintaining ultra-high $Q$, and they have the highest $Q/V$ metrics reported to date \cite{UltraSlot,ShurenArxiv}. Hence, PCNs are one of the most promising nanophotonic building blocks for realizing on-chip photonic devices with improved performance metrics, lower power consumption, and smaller footprint. 

Previous work on PCN cavities has shown these devices may be resonantly excited in various coupling configurations. Out of plane excitation methods such as the resonant scattering technique have proven useful for measuring exceptionally high quality factors \cite{2DfanoHighQ,HighQoutofPlanenanobeam}, leveraging optical forces \cite{slicedphc1,slicedphc2} and sensing changes to environmental refractive index \cite{2Dsensingslot}. While out-of-plane functionality is useful for many applications, in-plane functionality is a requirement for on-chip integration of PCN cavities with other photonic components. One of the most widespread and well-understood configurations for PCNs with in-plane functionality is in-line coupling, where light is coupled to resonant PCN modes from the edge of the cavity \cite{OpMechHighQ/V,UltraSlot,JudsonSlot,ShurenArxiv}. This configuration produces spectral peaks at resonant wavelengths, low transmission at non-resonant wavelengths in the PCN band-gap, and high transmission outside the band-gap. The importance of sharp, high-contrast spectral features for signal processing applications in combination with the relatively small device footprint of PCN cavities makes them an attractive element for dense, on-chip optical circuits. While in-line coupling to PCN cavities has demonstrated promise for on-chip signal processing, utilization of in-line coupling in compact multiplexing schemes (e.g. serial integration and add/drop filtering) is generally not feasible. This is due to low baseline transmission and limited input and output ports in an in-line coupling configuration. Multiplexing functionality can be greatly improved by evanescently side-coupling to PCN cavities \cite{EricksonSensorArrays,MultiplexGasSensor}. 

%In this configuration, non-resonant light passes with high transmission through a bus waveguide lateral to a PCN cavity. When light of a frequency resonant to the PCN cavity passes by the cavity, however, it couples into the cavity and is not transmitted through the bus waveguide, causing a sharp dip in transmission. 

Side-coupling to PCN cavities has not been as intensively studied for design principles as the in-line coupling configuration. Prior work implementing simple bus waveguides for side-coupling into PCN cavities has briefly mentioned basic design concepts without detailed study \cite{CoupledNanobeamCavities,Sidecoupled-plain}. We note that design analysis of PCN cavities laterally coupled to a waveguide Fabry-Perot resonator \cite{SidecoupledFano1,SidecoupledFano2,SidecoupledFano3}, waveguides laterally coupled to ring resonators \cite{RingResCouplingYariv_ElecLett2000}, fibers coupled to phase matched photonic crystal waveguides \cite{CoupledtoOpticFiber}, and generalized optical coupling formalisms \cite{YarivPaper,PhotonicCrystalsBook} have all been discussed in depth. Evanescent coupling to PCN cavities has also been discussed with greater detail in relation to lasing applications \cite{RaineriPaper1,RaineriPaper3_OptLett2016}. In this work, we show through simulation and experiment how tuning the overlap in state-space between a simple bus waveguide mode and PCN cavity resonances can be carried out as a straightforward approach to efficiently excite high $Q$ PCN resonances.

\section{Design Considerations}
\label{sec:Design}
\subsection{Evanescent Coupling and State-Space Overlap}
Evanescent coupling between two optical devices, such as waveguides and resonant cavities, occurs when energy is transferred from a photonic mode in one device to a photonic mode in the other device by means of the evanescent tails of the electric field distributions. The rate of energy exchange between a resonant cavity and its environment can be characterized by two connected metrics: the quality factor, $Q$, and the cavity field decay rate, $\Gamma$. In this paper, both of these terms will be used. The relationship between $Q$ and $\Gamma$ is
\begin{equation}
Q \propto \frac{1}{\Gamma} ,
\label{eqn:QandGamma}
\end{equation}
as $Q$ is proportional to the average photon lifetime in the cavity and $\Gamma$ is inversely proportional to the lifetime of the cavity. Both $Q$ and $\Gamma$ can be used to describe the overall losses of the cavity or particular categories of cavity loss, such as losses due to resonator coupling ($Q_{c}, \Gamma^{c}$) or intrinsic resonator losses ($Q_{0}, \Gamma^{0}$).

The efficacy of evanescent coupling, or extent of energy exchange between optical devices as defined by the coupling decay rate, $\Gamma^{c}$, is dependent on the overlap of the two modes in state-space (e.g., physical space, k-space, $\omega$-space). In physical space, the overlap of the modes is controlled by adjusting the size of the physical gap separating the evanescently coupled optical devices. In k-space, the modal overlap may be adjusted by offsetting the wave-vectors supported between the two devices. In the case of an optical device with discontinuous values allowed in state-space, such as a resonator where limited combinations of optical frequencies and wave-vectors are supported, coupling between the resonant device and a feeding device, such as a waveguide, may only occur when the mode of the feeding device is similar to an allowed mode in the resonator. Considering one of the most common evanescently coupled photonic components, a ring resonator, coupling between a straight bus waveguide and a resonant ring occurs when the bus waveguide mode is of the same frequency as the resonant mode in the ring. Generally, the wave-vectors of the bus waveguide mode and resonant mode in the ring are very similar in the coupling region when the bus and ring waveguides have similar widths. Consequently, the rate of energy exchange between the bus waveguide and resonant ring is almost entirely determined by the gap spacing between the bus waveguide and the ring. Hence, to approach critical coupling for ring resonator devices, usually a small parameter sweep of coupling gap sizes, in experiment or simulation, is sufficient to realize an optimal device configuration. 

For evanescent coupling between a bus waveguide and a PCN cavity, however, it cannot be assumed that the wave-vectors of the bus waveguide and PCN cavity are similar. Additionally, the optical coupling efficiency, $\eta$, in evanescently coupled PCN cavities and ring resonators is described by different solutions of the scattering-theory of waveguide-resonator coupling \cite{YarivPaper}. Different solutions for $\eta$ exist between ring resonators and PCN cavities because ring resonator modes are degenerate, two distinct modes exist at the same frequency, while PCN cavity modes are non-degenerate and only one resonance mode exists at a particular frequency. PCN modes are generally non-degenerate unless otherwise engineered to display degenerate behaviour \cite{CoupledNanobeamCavities,YarivPaper}. Therefore, additional design considerations are necessary to maximize $\eta$ through properly balancing the intrinsic losses, $Q_{0}$ and $\Gamma^{0}$, and the coupling losses, $Q_{c}$ and $\Gamma^{c}$, of the cavity. This may be achieved by ensuring that the modes of the bus waveguide and PCN cavity have the appropriate overlap in both physical and k-space.

\subsection{Calculating Wave-Vectors of Bus Waveguide and PCN Cavity}
\label{subsection:statespaceoverlap}
As a first step toward tuning the degree of k-space overlap between a PCN cavity mode and bus waveguide mode, we define the wave-vectors of these modes. Assuming the PCN cavity mode is designed to be at the edge of the Brillouin zone, the wave-vector of the resonant cavity mode is given by
\begin{equation}
k_{cav} = \frac{0.5}{a} \left(\mathrm{\frac{1}{nm}}\right),
\label{eqn:kcav}
\end{equation}
where $a$ is the period of the air holes comprising the PCN. In this work, the PCN cavity mode is a dielectric mode. The band structure of the PCN, which depends on the dimensions of the unit cell, determines the dielectric band edge frequency and, thus, the dielectric cavity mode resonance frequency, as discussed in \textbf{Section \ref{sec:Simulations}}. In order to couple light from a bus waveguide to the PCN cavity mode, there must exist a bus waveguide mode that has the same frequency as the PCN cavity mode. The wave-vector of such a bus waveguide mode is given by
\begin{equation}
k_{wg} = n_{eff-wg}\cdot\frac{1}{\lambda_{res}} \left(\mathrm{\frac{1}{nm}}\right),
\label{eqn:wvgeff}
\end{equation}
where $n_{eff-wvg}$ and $\lambda_{res}$ are the effective index of the propagating waveguide mode and the free-space wavelength corresponding to the PCN cavity resonance frequency, respectively. The condition necessary for maximum overlap in k-space, or phase matching, between the bus waveguide and PCN cavity modes can be obtained by setting $k_{cav} = k_{wg}$, as shown in Eqn. (\ref{eqn:neff}). 
\begin{equation}
n_{eff-wvg}= \frac{0.5}{a}\cdot\lambda_{res}
\label{eqn:neff}
\end{equation}
Therefore, in order to control the rate of energy exchange rate, $\Gamma^{c}$, between a side-coupled bus waveguide and PCN cavity, we can adjust \emph{(1) the coupling gap between the bus waveguide and the cavity} as well as \emph{(2) the offset between the k-vector of the cavity and the k-vector of the bus waveguide}. By tuning these parameters, $\Gamma^{c}$ can readily be altered. With the proper choice of $\Gamma^{c}$ with respect to $\Gamma^{0}$, we can realize large $\eta$ for resonance modes, which is discussed in the next section. Additionally, a computationally efficient method to quickly determine a functional combination of bus waveguide width and coupling gap for practical devices is presented in \textbf{Section \ref{section:experimental}}.

\subsection{Coupling Efficiency for In-Line vs Side-Coupled Devices}
\label{subsection:coupling comparison}
\begin{figure}[b]
\center
\includegraphics[scale = .43,trim={1.6cm 0cm 2.5cm 0cm},clip,page=1]{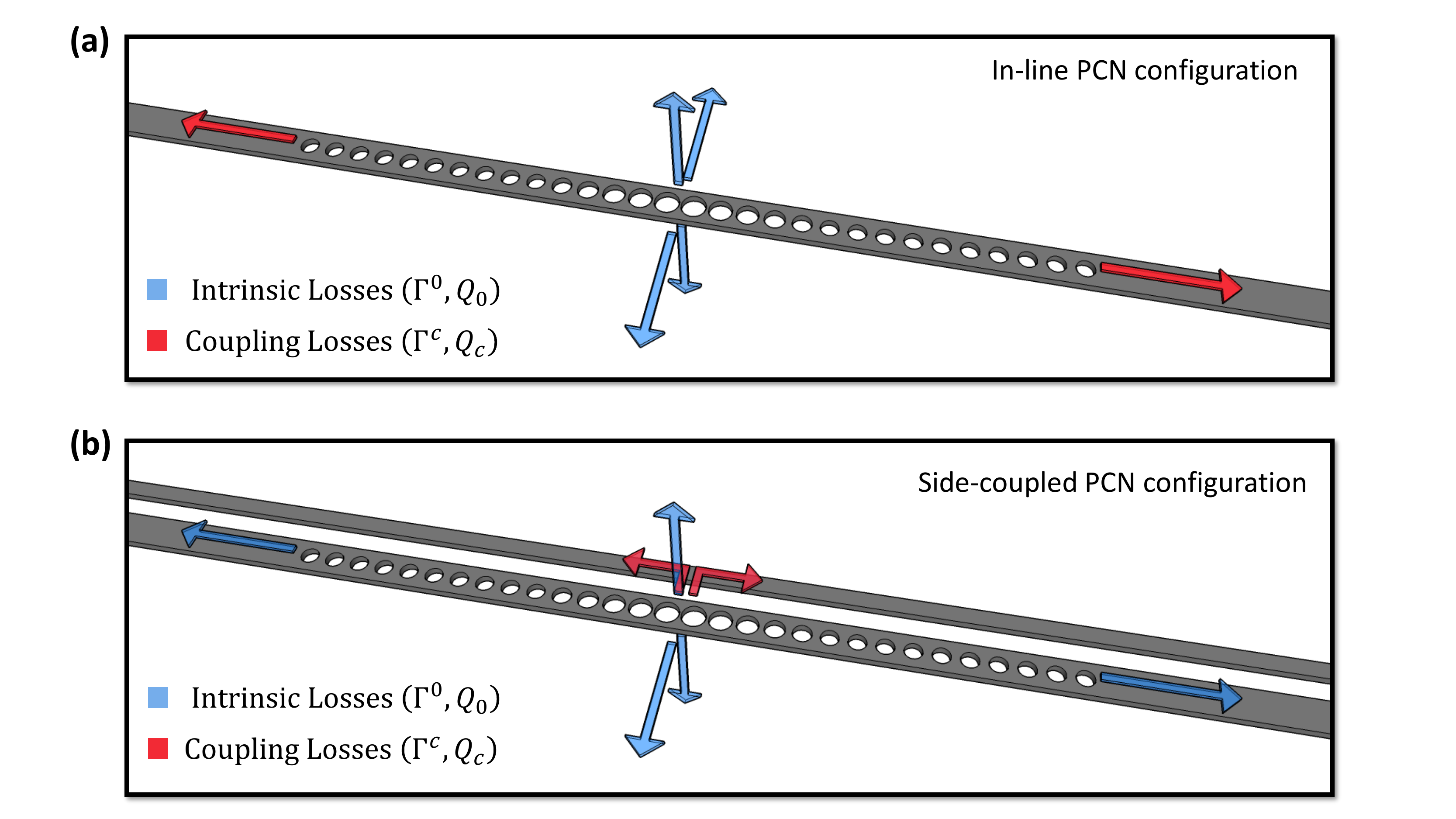}
\caption{Schematic illustration of coupling losses associated with (a) in-line and (b) side-coupled PCN cavities. a) Intrinsic losses (blue) correspond to resonant fields decaying into modes which scatter into free-space surrounding the device. Coupling losses (red) correspond to resonant fields decaying into modes guided by the waveguides present at the edges of the cavity. b) Intrinsic losses (blue) correspond to resonant fields decaying into modes which scatter into free-space surrounding the device and through the edges of the cavity. Coupling losses (red) correspond to resonant fields decaying into modes guided by the lateral bus waveguide.} 
\label{fig:Losses}
\end{figure}

In discussing the design of side-coupled PCN cavities, it is important to outline differences in coupling which alter optimal PCN geometric parameters for use in a side-coupled configuration from those for use in an in-line configuration. For in-line coupling to PCN cavities, the resonant coupling efficiency of the PCN cavity to the coupled waveguides, $\eta$, and transmission coefficient, $T$, on resonance can be calculated by the equation
\begin{equation}
\eta_{(in-line)} = T_{(in-line)} = \frac{\left(\Gamma^{c}\right)^2}{\left(\Gamma^{0}+\Gamma^{c}\right)^2} \:,
\label{eqn:etainline}
\end{equation}
where $\Gamma^{c}$ is the coupling decay rate and $\Gamma^{0}$ is the intrinsic decay rate of the cavity \cite{YarivPaper}. In the in-line case, $\Gamma^{c}$ refers to of the decay rate of the resonance field into the coupling waveguides at the edges of the cavity \cite{YarivPaper}. $\Gamma^{0}$ represents the decay rate of the resonance field that is not coupled into the waveguides present at the edges of the cavity. The loss associated with $\Gamma^{0}$ is coupled into free-space or the underlying substrate (if present) through scattering. These losses are schematically described in \textbf{Fig. \ref{fig:Losses}(a)}. When designing cavities for in-line operation with transmission peaks on resonance, it is important to note that $\Gamma^{0}$ and $\Gamma^{c}$ are, in general, entangled quantities. By increasing the length of tapering or the number of mirror pairs in a PCN cavity, it is possible to significantly decrease the scattering decay rate, $\Gamma^{0}$, as longer taper lengths and extra mirror pairs more gradually and strongly confine the resonance modes. Longer taper lengths and mirror pairs, however, also work to decrease the rate of energy exchange with the feeding waveguide, $\Gamma^{c}$, as the mode becomes less coupled to the surrounding waveguides by a spatially larger potential barrier induced by added taper and mirror segments. Previously reported results on in-line devices show that the competing rates of increase between $\Gamma^{0}$ and $\Gamma^{c}$ can lead to decrease in on-resonance transmission and resonance coupling efficiency for in-line PCN cavities with higher $Q$ \cite{BigDeterministicDesign}. Thus a proper compromise between $Q$ and $\eta$ must be made for any in-line application.

In side-coupled devices, resonance conditions cause dips in transmission. The equations for transmission and coupling efficiency defined by low resonance transmission are
\begin{equation}
T_{(side-coupled)} = \frac{\left(\Gamma^{0}\right)^2}{\left(\Gamma^{0}+\Gamma^{c}\right)^2}
\label{eqn:Tsidecoupled}
\end{equation}
\begin{equation}
\eta_{(side-coupled)} = 1-T_{(side-coupled)} = \frac{2\left(\frac{\Gamma^{c}}{\Gamma^{0}}\right)+\left(\frac{\Gamma^{c}}{\Gamma^{0}}\right)^2}{\left(1+\frac{\Gamma^{c}}{\Gamma^{0}}\right)^2}.
\label{eqn:etasidecoupled}
\end{equation}
Equation (\ref{eqn:etasidecoupled}) is written as a function of $\Gamma^{c}/\Gamma^{0}$ for utility when calculating theoretical transmission from simulations, which will be discussed further in \textbf{Section \ref{subsec:Estimation}}. \emph{It is important to note that $\Gamma^{0}$ and $\Gamma^{c}$ refer to different decay mechanisms in side-coupled and in-line configurations} \cite{YarivPaper}. These differences are illustrated in \textbf{Fig. \ref{fig:Losses}}. For side-coupling, $\Gamma^{c}$ refers to the coupling decay rate of the resonance field into the side-coupled bus waveguide and $\Gamma^{0}$ refers to the decay rate associated with the resonance field decaying both through free-space scattering and through the edges of the cavity. The utilization of a side-coupler allows decoupling of intrinsic and coupling losses to the PCN cavity. So long as the introduction of the bus waveguide is realized in a way to not significantly perturb the cavity mode, the intrinsic losses result solely from the cavity geometry, whereas the coupling losses are determined by the overlap of the cavity and bus waveguide modes in physical and k-space. The decoupling of intrinsic and coupling losses could potentially enable larger coupling efficiency, $\eta$, at high quality factors than what is possible in an in-line configuration. The loaded $Q$ of a system with high $\eta$ is then determined by the ratio of the decay rates, $\nicefrac{\Gamma_{c}}{\Gamma_{0}}$. As described in \textbf{Section \ref{subsection:statespaceoverlap}}, $\Gamma^{c}$ may be directly controlled by tuning the overlap of the two optical modes in state-space. $\Gamma^{0}$ may be controlled by altering the cavity taper length and number of extra mirrors at the edges of the cavity.

\section{Simulations}
\label{sec:Simulations}
\subsection{Simulation Specifications and PCN Design}
Two-dimensional (2D) finite-difference time-domain (FDTD) simulations were carried out in Lumerical to design and characterize side-coupling effects on PCN cavities. Two-dimensional analysis was chosen due to the significant computational resources required to simulate parameter sweeps for high Q cavities. Because two dimensional simulations allow for separation of coupled optical modes in both physical and k-space, they allow sufficient degrees of freedom to probe effects of proximity in state-space on coupling and performance of PCN cavities. These degrees of freedom consist of (1) \emph{the spatial dimension along which the bus waveguide and PCN cavity are separated} and (2) \emph{the dimension in k-space by which the two optical modes are offset}. While the exact distributions of optical modes in physical and k-space may differ between 2D and 3D simulations along the axes of separation, the trends in PCN cavity resonance depth as a function of bus waveguide width and coupling gap size are expected to be similar for 2D and 3D analysis because of the similar degrees of freedom utilized in both cases. It is also important to note that the coupling formalism used here, which is detailed in \cite{YarivPaper}, is independent of device dimensionality. 

\begin{figure}[t]
\center
\includegraphics[scale = .43,trim={2cm 2.5cm 1.5cm 3.3cm},clip,page=2]{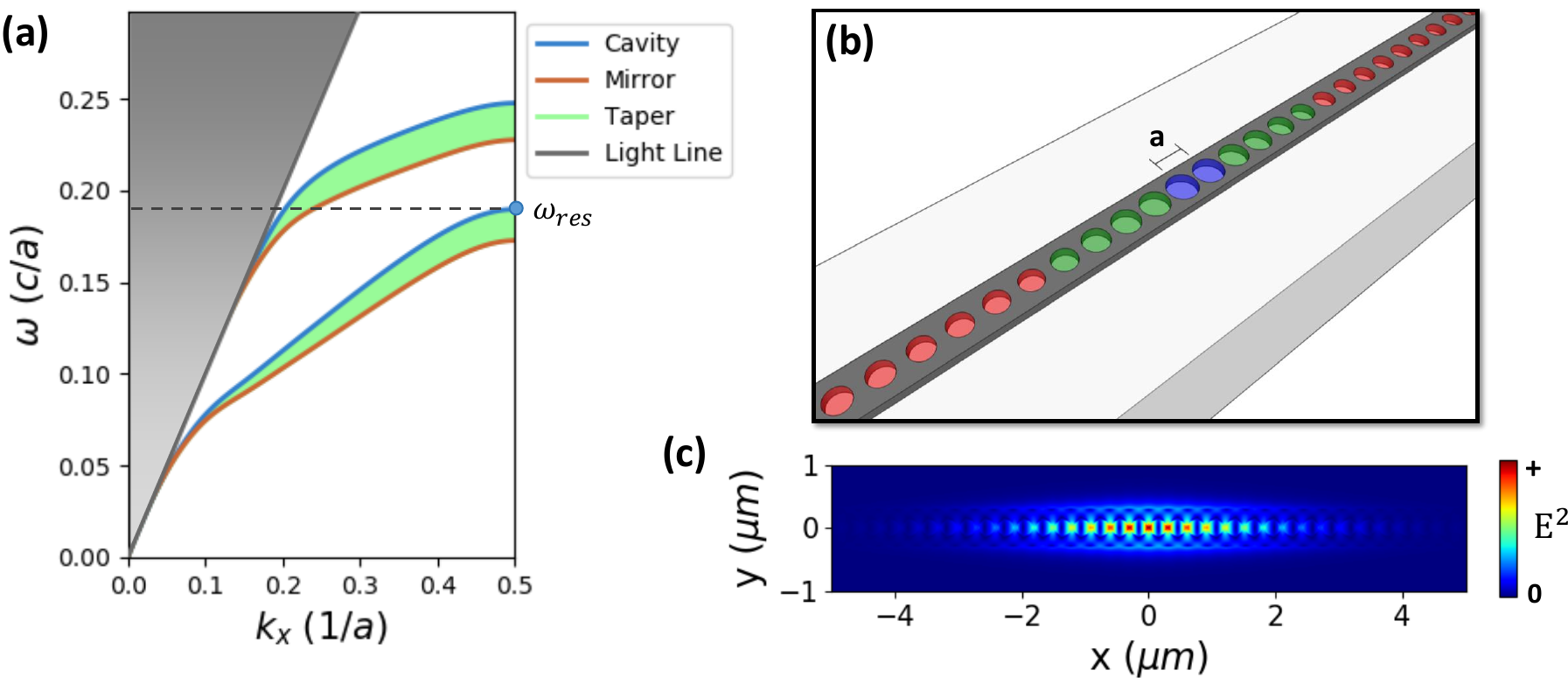}
\caption{a) Band diagram of TE modes associated with end mirror (red), taper (green), and cavity (blue) regions of the photonic crystal cavity shown in b). These bands were calculated via 2D FDTD simulations. b) Schematic illustration of deterministically designed PCN with cavity resonance state within the band gap of the end mirror segments. c) Mode profile of fundamental resonance of designed PCN cavity.} 
\label{fig:bandstruc}
\end{figure}

The deterministic design approach for PCN cavities was followed in order to reduce scattering losses and increase the intrinsic quality factor of the PCN cavities \cite{DeterministicDesign,BigDeterministicDesign}. The specific dimensions utilized for the PCN cavities are as follows. The width of the nanobeam was selected to be $nw = 600 \mathrm{nm}$ and the period was set to $a = 300 \mathrm{nm}$ with deterministic hole tapering from a radius of $103 \mathrm{nm}$ in the mirror segments to a radius of $130 \mathrm{nm}$ in the cavity region. We chose $nw = 600 \mathrm{nm}$ in the 2D simulations to tailor the lateral confinement of the optical mode to better mimic the performance of 3D devices. The modes of fabricated PCN devices (i.e. 3D devices with finite thickness), such as those studied in \textbf{Section \ref{section:experimental}}, are less well confined compared to those of a 2D PCN device in simulation with the same $nw$. Using a smaller $nw$ in 2D simulations helps to maintain a similar level of lateral field confinement. To maintain the cavity resonance frequency near 1550nm with this $nw$, a relatively short period of $a=300$nm was employed. A taper length of 20 unit cells was utilized along with 10 identical unit cells in the end mirror regions to yield a well-confined resonance mode. The band structure of the unit cells comprising the PCN and a schematic of the PCN are shown in \textbf{Fig. \ref{fig:bandstruc}(a,b)}. By gradually modifying the unit cell band structure from the ends of the mirror region to the cavity center, a locally allowed state is created at the edge of the Brillouin zone. \textbf{Figure \ref{fig:bandstruc}(a)} shows that in a PCN designed for a dielectric cavity resonance mode, the resonance frequency is determined by the dielectric band edge frequency for unit cells in the cavity center, where the cavity resonance wave-vector is $k_{x}=\nicefrac{0.5}{a}$ (i.e., at the edge of the Brillouin zone). The profile of the fundamental cavity resonance mode of the designed PCN cavity is shown in \textbf{Fig. \ref{fig:bandstruc}(c)} and has a simulated $Q_{0}=1.7\times10^{8}$. This calculated quality factor is the intrinsic quality factor and is associated with the intrinsic losses, $\Gamma^0$, due to the absence of a side-coupler in these simulations.

\subsection{Simulating Modal Distribution of Wave-Vectors}
\begin{figure}[t]
\center
\includegraphics[scale = .39,trim={0cm 2.5cm 0cm 2.5cm},clip,page=3]{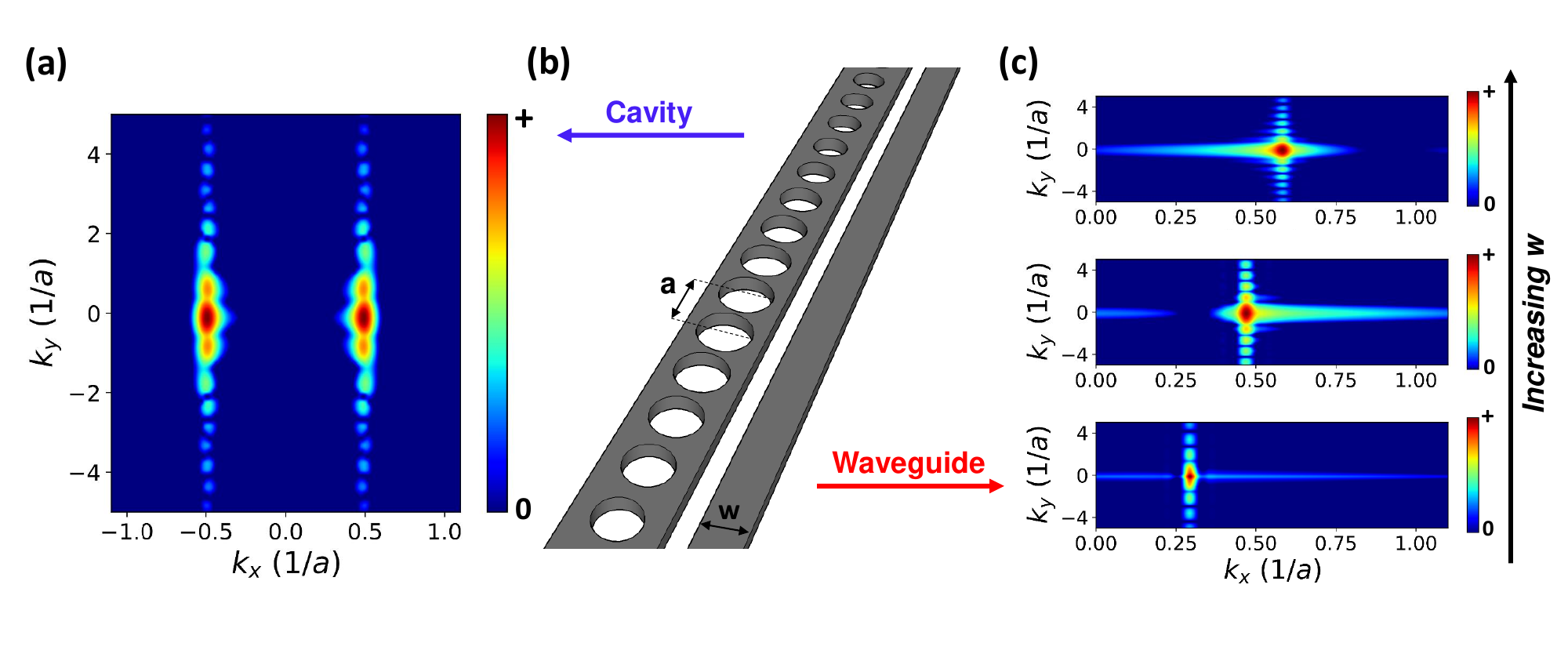}
\caption{a) k-space plot of optical cavity resonance taken from the Fourier transform of the field profile shown in \textbf{Fig. \ref{fig:bandstruc}(c)}. b) Schematic illustration of side-coupling configuration for PCN cavity. c) k-space plots of bus waveguide modes of various waveguide widths ($w$ = 250nm, 290nm, and 400nm). By increasing the waveguide width, it is possible to increase the k-vector of the waveguide mode due to an increase in the modal effective index.}
\label{fig:wavevector}
\end{figure}

To explore the properties of evanescent coupling discussed in \textbf{Section \ref{sec:Design}}, we carried out simulations to examine how tuning the overlap in k-space between modes of a straight bus waveguide and a PCN cavity affects side-coupled performance. We first examined the Fourier transform of the electric field profile of the PCN cavity fundamental mode (\textbf{Fig. \ref{fig:bandstruc}(c)}) to visualize the cavity resonance profile in k-space, as shown in \textbf{Fig. \ref{fig:wavevector}(a)}. The wave-vectors are localized at $|k_{x}| = \nicefrac{0.5}{a}$, as expected for a deterministically designed cavity. Next, in order to tune the wave-vector of light evanescently coupled from the bus waveguide, the effective index of the bus waveguide was altered by adjusting $w$, the width of the bus waveguide. Lumerical MODE Solutions was used to calculate the effective index of the fundamental TE-mode in the bus waveguide for different values of $w$. Increasing $w$ leads to an increase in the wave-vector of the guided mode. \textbf{Fig. \ref{fig:wavevector}(c)} shows the Fourier transform of the electric field distribution in the bus waveguide for three different waveguide widths, $w$ = 250nm, 290nm, and 400nm, corresponding to k-space distributions centered around wave-vectors that are less than, nearly equal to, and greater than the wave-vector of the resonant PCN cavity mode, respectively. FDTD simulations of the waveguide mode profile in k-space were done with a 5000fs pulsed mode source at the resonant wavelength of the PCN cavity and measured by an electric field monitor extending through the waveguide. The long mode source was utilized to minimize transient effects and to ensure guided optical fields extended throughout the full waveguide region monitored.

\subsection{Computational Results of State-Space Overlap on Coupling}
\label{subsection:computationalresults}
\begin{figure}[t]
\center
\includegraphics[scale = .40,trim={0.3cm 2.5cm 0.3cm 2.5cm},clip,page=4]{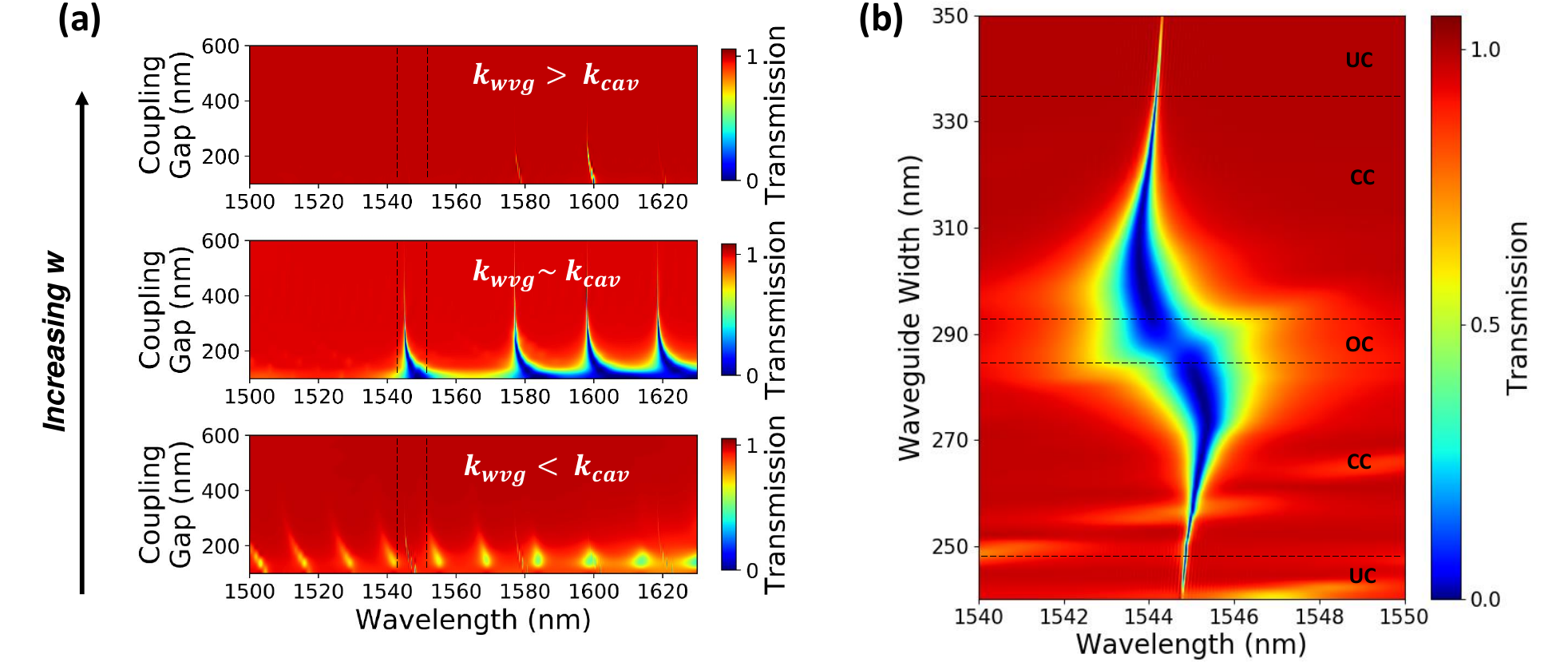}
\caption{a) Simulated transmission from side-coupled PCN cavities with bus waveguide widths of 250nm (Bottom), 290nm (Middle) and 400nm (Top) and different coupling gap sizes. b) Simulated transmission from side-coupled PCN cavities with fixed coupling gap ($g$ = 150nm) and various waveguide widths. Under-coupled, over-coupled and critically coupled regions are denoted by \textbf{UC}, \textbf{OC} and \textbf{CC}, respectively.}
\label{fig:Concept}
\end{figure}

As stated previously, controlling the overlap of the bus waveguide and cavity resonance modes in both k-space and physical space allows control of the rate of energy exchange due to coupling, $\Gamma^{c}$. To investigate the effect of modal overlap in state-space on the efficacy of resonance excitation in PCN cavities, 2D calculations in Lumerical FDTD Solutions were carried out. These simulations calculated the transmission spectrum of the PCN specified at the beginning of \textbf{Section \ref{sec:Simulations}} when side-coupled using different bus waveguide and coupling gap dimensions. As discussed previously, we believe the results of the less computationally intensive 2D simulations are able to sufficiently illustrate trends in the physical behavior of the side-coupled PCN system, and we show this to be the case in comparison to experimental results in \textbf{Section \ref{section:experimental}}. The three aforementioned bus waveguide widths were considered and the coupling gap size was varied between 100 $ - $ 600nm. The resulting transmission spectra of these simulations are shown in \textbf{Fig. \ref{fig:Concept}(a)}. 

The most pronounced resonances occur when the wave-vector of the bus waveguide mode closely matches that of the PCN cavity resonance (i.e., $w$ = 290nm). For the side-coupled PCN with the narrowest bus waveguide width, the transmission features that appear for small coupling gap size are most likely Fabry-Perot fringes that result from the finite size of the bus waveguide used in the simulation and the large evanescent tail of the electric field that extends from the narrow waveguide. Two narrow, but rather shallow resonances appear in the transmission spectra of the side-coupled PCN with the widest bus waveguide width and smaller coupling gaps. Overall, the data in \textbf{Fig. \ref{fig:Concept}(a)} suggest that the coupling decay rate, $\Gamma^{c}$, between the PCN cavity and bus waveguide modes is strongly affected by the wave-vector mismatch between the bus waveguide and cavity resonance modes. 

It is important to note that in the case where the waveguide width, $w$, is narrowest ($w$ = 250nm), the evanescent tail of the electric field extending from the bus waveguide is the largest due the mode being more delocalized and having a lower effective index. However, despite the fact that the 250nm wide bus waveguide has the largest field overlap in physical space with the PCN cavity, coupling from the bus waveguide of width $w$ = 290nm is significantly improved. Although the spatial field overlap between the bus waveguide with $w$ = 290nm and the PCN cavity is not the largest of the three bus waveguides considered, the wave-vector of the bus waveguide is the most closely matched to that of the PCN cavity resonance. Therefore, we conclude that the effect of modal overlap in k-space may, at times, dominate the effect of field overlap between devices in physical space.

Next, we simulated transmission spectra for side-coupled PCNs with constant $g$ = 150nm and variable $w$ between 240nm $ - $ 350nm to further explore the dependence of the k-space overlap on the energy exchange rate between the bus waveguide and PCN cavity. \textbf{Figure \ref{fig:Concept}(b)} shows that the width and depth of the fundamental transmission resonance is highly dependent on the bus waveguide width, and, hence, its supported wave-vector. We note that the slight change in resonance wavelength as a function of $w$ is likely due to the different levels of interaction between the bus waveguide mode and the PCN cavity resonance mode, which affects the cavity resonance frequency. By increasing $w$, at a constant $g$, the modal overlap between the bus waveguide and PCN cavity is altered in k-space and the coupling regime is altered. 

\subsection{Coupling Regimes in Side-Coupled PCN Cavities}
To categorize the effectiveness of a side-coupled geometry, it is important to define coupling regimes which describe the device performance. In \textbf{Fig. \ref{fig:Concept}(b)}, we label different side-coupled geometries as under-, over- and critically-coupled, borrowing the common terminology used to identify different coupling regimes for ring resonators. While the transmission spectra of evanescently coupled photonic crystals and ring resonators are not described by the same equations, similar physical features can be observed in each case. 
%\emph{To better identify regions in parameter space with differing side-coupled PCN performance, we use these physically observable effects to define coupling regimes}.

The conventional use of the terms over-coupled, under-coupled, and critically coupled in ring resonators can be summarized as follows:
\begin{myindentpar}{0cm}
\textbf{(Ring, 1)} Critically coupled ring resonators have $\eta = 1$, meaning $T = 0$ on resonance.  \\
\textbf{(Ring, 2)} Under-coupled ring resonators have $\eta < 1$ and higher $Q$ than the critically. coupled case\\
\textbf{(Ring, 3)} Over-coupled ring resonators have $\eta < 1$ and lower $Q$ than the critically coupled case.
\end{myindentpar}
In this work we utilize the terms over-coupled [\emph{o.c.}], under-coupled [\emph{u.c.}], and (near) critically coupled [(\emph{n.})\emph{c.c}]. Here, critical coupling in evanescently coupled PCN cavities is defined by the following criteria:
\begin{myindentpar}{0cm}
\textbf{(PCN, 1)} Critically-coupled PCN cavities are characterized by $\eta \approx 1$, meaning $T \approx 0$ on resonance. \\
\textbf{(PCN, 2)} If critical coupling cannot be achieved in a particular parameter space, we utilize the term near-critical coupling to define the local maximum in $\eta$ observed in that parameter space, $\eta_{max} \le 1$. This definition is particularly relevant to results in \textbf{Section \ref{section:experimental}}.
\end{myindentpar}
Under-coupled and over-coupled regimes for evanescently coupled PCN cavities are identified based on the $Q_{total}$ and $\eta$ of the side-coupled PCN cavity. $Q_{total}$ and $\eta$ can be directly determined from the transmission spectrum of the side-coupled PCN cavity. The over-coupled and under-coupled regimes are defined as:
\begin{myindentpar}{0cm}
\textbf{(PCN, 3)} Under-coupled cavities have narrow, shallow resonance dips ($Q_{u.c.} > Q_{c.c.}$ and $\eta_{u.c.} < \eta_{max}$). \\
\textbf{(PCN, 4)} Over-coupled cavities have broad, shallow resonance dips ($Q_{o.c.} < Q_{c.c.}$ and $\eta_{o.c.} < \eta_{max}$).
\end{myindentpar}
Utilizing Eqn. (\ref{eqn:etasidecoupled}), we directly connect the coupling regimes to the ratio of $\nicefrac{\Gamma^{c}}{\Gamma^{0}}$ in the system. 
\begin{myindentpar}{0cm}
\textbf{(PCN, 5)} Critical coupling occurs when $\Gamma^{c} \gg \Gamma^{0}$, meaning $\nicefrac{\Gamma^{c}}{\Gamma^{0}} \gg 1$. \\
\textbf{(PCN, 6)} Over-coupling and under-coupling occur when $\Gamma^{c} \lesssim \Gamma^{0}$, meaning $\nicefrac{\Gamma^{c}}{\Gamma^{0}} \lesssim 1$. 
\end{myindentpar}
With the criterion $\Gamma^{c} \gg \Gamma^{0}$, critical coupling for side-coupled PCN cavities occurs over a range of $\nicefrac{\Gamma^{c}}{\Gamma^{0}}$. This allows critical coupling to occur over a range of $w$ values instead of at a single $w$ as shown in \textbf{Fig. \ref{fig:Concept}(b)}. The distinction between over-coupling and under-coupling PCN cavities lies in the value of $\Gamma^{0}$ and consequently $Q_{0}$. In the under-coupling and critically coupled regime, $\Gamma^{0}$ and $Q_{0}$ do not change significantly. In the over-coupled regime, $\Gamma^{0}$ is larger and $Q_{0}$ is smaller than what is observed in the critically coupled and under-coupled case. The heightened intrinsic losses of the cavity when over-coupled are due to the large modal overlap between the PCN cavity and the feeding bus waveguide. The increased interaction in the over-coupled regime significantly increases unguided scattering from the cavity resonance and consequently reduces both $\eta$ and $Q_{total}$ from the critically coupled case. Over-coupling phenomena in PCN cavities has also been observed in \cite{RaineriPaper3_OptLett2016}.

We note the over-coupled region in \textbf{Fig. \ref{fig:Concept}(b)} is centered around $w \sim$ 290nm, near where data from \textbf{Fig. \ref{fig:wavevector}(c)} suggests $k_{cav} = k_{wg}$. This observation suggests that phase matching to PCN cavities may disproportionately increase $\Gamma^{0}$ compared to $\Gamma^{c}$, reducing the ratio $\nicefrac{\Gamma^{0}}{\Gamma^{c}}$ and increasing the potential to over-couple the cavity. 

The over-coupling phenomenon results from the introduction of the bus waveguide increasing $\Gamma^{0}$. Solutions to avoiding this phenomenon include (1) \emph{offsetting the two optical mode distributions in k-space } and (2) \emph{ensuring a large enough coupling gap is used to control the evanescent field overlap between the cavity and bus waveguide}. In the cases probed through simulations and experiments in this paper, we show optimal configurations using non phase-matched conditions, but PCN cavities with phase-matched conditions could potentially operate with high $\eta$ if a sufficient coupling gap is implemented to properly balance $\Gamma^{c}$ and $\Gamma^{0}$.

\section{Experiments}
\label{section:experimental}
\subsection{Photonic Crystal Nanobeam Cavity Design}
\label{subsection:devicedesign}
As mentioned in \textbf{Section \ref{subsection:coupling comparison}}, it is desirable to use a PCN cavity with high $Q_{0}$ for efficient side-coupling. Accordingly, we implemented a PCN design with sufficient taper length and mirror strength to realize a simulated $Q_{0} \sim 10^{6}$. The fabricated PCN design had a target width of $nw = 700 \mathrm{nm}$ and period of $a = 350 \mathrm{nm}$ with deterministic hole tapering from a radius of $105 \mathrm{nm}$ in the mirror segments to a radius of $127 \mathrm{nm}$ in the cavity region. The taper regions on either side of the cavity center consisted of 20 unit cells each, and the mirror regions consisted of 10 unit cells each. 

Differences in the bus waveguide and PCN dimensions between the 2D FDTD simulations (\textbf{Section \ref{sec:Simulations}}) and fabricated structures were required to compensate for the different mode confinement in a 2D simulation compared to a fabricated (i.e., 3D) structure. As stated earlier, the PCN width was decreased in simulation to $nw$ = 600nm to better mimic the modal confinement in a fabricated PCN cavity with a traditional $nw$ = 700nm width. The period $a$ and radii of the PCN holes were accordingly adjusted in the fabricated PCN cavities to follow the deterministic design approach for a PCN with $nw$ = 700nm in a 220nm thick SOI device layer \cite{DeterministicDesign,BigDeterministicDesign}. 

\subsection{Method for Testing Effect of Wave-Vector Overlap}
The sensitivity of device performance to variation in $k_{wvg}$ demonstrated in \textbf{Section \ref{sec:Simulations}} suggested the most practical way to measure the effects of k-vector mismatch and fabricate functional devices is to sweep the feeding bus waveguide width while maintaining a constant coupling gap size and PCN cavity geometry. Conceptually, this means we are holding $k_{cav}$ constant while varying $k_{wvg}$. In this way, our experiments are able to tolerate deviations in the PCN cavity resonance conditions that arise from fabrication variation by having several side-coupled devices, each with a bus waveguide supporting a mode with a slightly different k-vector. We note that this type of empirical process optimization is commonly performed with resonant silicon photonic structures to account for device variation and additional loss factors that arise due to fabrication imperfections (e.g. roughness, size bias) that are not incorporated into simulations. We chose to sweep the waveguide width in a range around the $k_{cav} = k_{wvg}$ condition in order to experimentally observe all coupling regimes defined earlier. This fixed condition helps to constrain our side-coupler design and influences the choice of coupling gap in \textbf{Section \ref{subsec:Estimation}}.

\subsection{Side-Coupling Design Strategy}
\label{subsec:designstrategy}
In order to efficiently side-couple from a straight bus waveguide into a PCN cavity, $\nicefrac{\Gamma^{c}}{\Gamma^{0}} \gg 1$ must be realized to achieve critical coupling. As described in this section, a practical method for optimizing a side-coupling geometry to achieve this condition is the following:

\begin{myindentpar}{0cm}
(1) Simulate the PCN cavity without a side-coupled bus waveguide to obtain $Q_{0}$ \\
(2) Identify practical design constraints on the bus waveguide coupling gap and waveguide width \\
(3) Set either the waveguide width or coupling gap as a constant value\\
(4) Sweep the free parameter and compute $Q_{total}$ for each side-coupled system to determine $\nicefrac{\Gamma^{c}}{\Gamma^{0}}$ \\
(5) Choose a value of the free parameter which achieves large $\eta$ and sufficient $Q_{total}$ \\
\end{myindentpar}

Due to the manifestation of coupling and intrinsic losses in a side-coupled PCN, we can accurately estimate $Q_{0}$ in step (1) by simulating the cavity without the bus waveguide present. Step (1) can be used to verify that the PCN cavity design realizes high $Q_{0}$ for efficient side-coupling. 

In step (2), constraints such as the resolution of lithography and fabrication tolerances can limit the parameter space for sweeping device design. A result of this could be limiting the smallest $g$ used or the tolerance/resolution of $w$ and $k_{wvg}$ achievable. 

Making either $g$ or $w$ a constant value in step (3) can further limit the parameter space for optimization. A good choice for this constant value should be where $\Gamma^{c}$ is high compared to $\Gamma^{0}$. A small value of $g > 0$ compared to the evanescent tail length or a choice of $w$ to place $k_{wvg} \sim k_{cav}$ but $k_{wvg} \neq k_{cav}$ would be adequate for satisfying this condition. Experimental data in \textbf{Sections \ref{subsection:experimentaltransmission}} and \textbf{\ref{subsection:experimentallosses}} shows that a value of $k_{wvg}$ $1.5\%$ larger or smaller than $k_{cav}$ was sufficient to avoid over-coupling. Greater k-space offsets further reduce the chance of over-coupling the cavity.

By sweeping the free parameter in step (4), we can measure $Q_{total}$ for the resonance mode and determine $\nicefrac{\Gamma^{c}}{\Gamma^{0}}$ for each value of the free parameter. After the parameter sweep, each value of $\nicefrac{\Gamma^{c}}{\Gamma^{0}}$ can be utilized to calculate $\eta$ for the corresponding dimension of the free parameter. For the design of devices used in this paper, we sweep the coupling gap from 100nm to 500nm. The minimum coupling gap of 100nm was considered as it was slightly larger than what is feasible for fabrication. The maximum coupling gap of 500nm was considered as it is approximately one third of the wavelength of the fundamental resonance mode. The interaction region of an evanescent field may be approximated as one third of the wavelength in many cases as a rule of thumb. In experiment we alter the k-space overlap by sweeping the bus waveguide width through a range 100nm larger and smaller than the width which satisfies the $k_{wvg} = k_{cav}$ condition determined by MODE Solutions calculations. We, however, experimentally observe measurable resonance peaks within a $7\%$ offset range between $k_{wvg}$ and $k_{cav}$ which occurs within a waveguide range 50nm wider or narrower than the k-vector matched geometry. These results are discussed in greater detail in \textbf{Section \ref{subsec:Estimation}}. 

We note that a more precise and general way to simulate $\nicefrac{\Gamma^{c}}{\Gamma^{0}}$ for each side-coupled cavity can also be computed using similar methods to those in \cite{DeterministicDesign}. In \cite{DeterministicDesign}, the simulated flux and output modes at each field decay port are computed to directly extract $\nicefrac{\Gamma^{c}}{\Gamma^{0}}$ in each simulation without utilizing a separate calculation of $Q_{0}$. The method in \cite{DeterministicDesign} is not used here as it requires greater computational resources to implement than the method proposed in this work.

\subsection{Estimating Device Performance}
\label{subsec:Estimation}
We followed the design strategy outlined in \textbf{Section \ref{subsec:designstrategy}} to determine a coupling gap which yields high $\eta$ for $k_{cav} \sim k_{wvg}$. The discussed simulations considered the PCN cavity geometry detailed in \textbf{Section \ref{subsection:devicedesign}}. Simulations were completed via 3D FDTD methods using Lumerical FDTD. A non-uniform meshing parameter of 2 in Lumerical FDTD was used in these 3D calculations to reduce simulation resources. It is important to have the substrate present in these simulations if the substrate is present in experiment, as it can strongly affect both $\Gamma^{0}$ and $\Gamma^{c}$. 

The steps from our design strategy were executed as follows:
\begin{myindentpar}{0cm}
(1) We excited resonant modes in the PCN cavity with a TE dipole source in the center of the cavity and monitored the field decay rate. From this field decay rate, we calculated $Q_{0} = 1.0 \times 10^{6}$ for the fundamental PCN resonance mode without the bus waveguide present.\\
(2) Our simulation space was constrained to $g > 50 \mathrm{nm}$ and we specified a step size for $w$ no smaller than 10nm to be used in fabrication. \\
(3) The constraint $k_{wvg} \sim k_{cav}$ was set so we might use the platform to explore the effect of offsetting $k_{wvg}$ from $k_{cav}$ in experiment. This constraint led to a constant value of $w$ = 480nm used for the bus waveguide while sweeping $g$. In high resolution MODE Solutions calculations, $w$ = 480nm set $k_{wvg} = k_{cav}$. The difference in meshing between the MODE Solutions simulations and FDTD computations, however, created a non-zero offset between $k_{cav}$ and  $k_{wvg}$ in the FDTD simulations. The slight k-vector offset yielded $k_{cav} \sim k_{wvg}$ but $k_{cav} \neq k_{wvg}$. The k-vector offset between the high resolution MODE solutions simulations and low resolution FDTD simulations was large enough to avoid the over-coupling regime in these simulations, but using a more similar resolution between MODE solutions and Lumerical FDTD would lead to less of an offset between $k_{cav}$ and  $k_{wvg}$. We reiterate that experimental data in \textbf{Sections \ref{subsection:experimentaltransmission}} and \textbf{\ref{subsection:experimentallosses}} shows that a value of $k_{wvg}$ $1.5\%$ larger or smaller than $k_{cav}$ was sufficient to avoid over-coupling. For utilization of high resolution in both effective index calculations and $Q$ simulations, we recommend this minimum offset be observed. In the over-coupling regime, the approximation that $Q_{0}$ is constant does not apply. A more detailed method to calculate $\nicefrac{\Gamma^{c}}{\Gamma^{0}}$, such as a method similar to what is used in \cite{DeterministicDesign}, is necessary for estimating $\eta$ in over-coupled PCN cavities. \\
(4) A small, relatively quick ($\sim$ 1 - 2 hours per simulation), parameter sweep of $g$ was performed for $g = 100\mathrm{nm}$ to $g = 500\mathrm{nm}$ in 100nm increments. For each value of $g$, $Q_{total}$ was computed. To calculate $\nicefrac{\Gamma^{c}}{\Gamma^{0}}$ we utilized the equations
\end{myindentpar} 
\begin{equation}
\frac{1}{Q_{total}}=\frac{1}{Q_{0}}+\frac{1}{Q_{c}}+\Rightarrow Q_{c}=\frac{Q_{total}}{\left(1-\frac{Q_{total}}{Q_{0}}\right)}
\label{eqn:QtotToQcoupling}
\end{equation}
\begin{equation}
\frac{\Gamma^{c}}{\Gamma^{0}}=\frac{Q_{0}}{Q_{c}}=\frac{Q_{0}}{Q_{total}}-1 \:,
\label{eqn:GammaratiofromQratio}
\end{equation}

where Eqn. (\ref{eqn:QtotToQcoupling}) is re-arranged from \cite{DeterministicDesign} and Eqn. (\ref{eqn:GammaratiofromQratio}) follows from Eqn. (\ref{eqn:QandGamma}) with utilization of the result from Eqn. (\ref{eqn:QtotToQcoupling}). After extracting $Q_{total}$ for each simulated value and using the initially computed $Q_{0}$, we can easily find $\nicefrac{\Gamma^{c}}{\Gamma^{0}}$ for each value of the swept parameter. \\
(5) For the bus waveguide present at $g$ = 400nm with $w$ = 480nm we measured $Q_{total} = 7.5 \times 10^{4}$, which corresponds to a calculated value of $Q_{c}=8.1 \times 10^{4}$. Using the computed $Q_{total}$ and $Q_{0}$ we see $\nicefrac{Q_{0}}{Q_{total}} = 13.3$. Plugging this ratio into Eqn.(\ref{eqn:GammaratiofromQratio}), we see $\nicefrac{\Gamma^{c}}{\Gamma^{0}}$=12.3. With this value of $\nicefrac{\Gamma^{c}}{\Gamma^{0}}$, we can calculate the ideal coupling efficiency of this combination of bus waveguide width and coupling gap to be $\eta \sim 0.995$ using Eqn. (\ref{eqn:etasidecoupled}), fulfilling the requirements for critical coupling. 

An important part of the decision to use a high ratio of $\nicefrac{\Gamma^{c}}{\Gamma^{0}}$ in designing side-coupled cavities for fabrication is noting that fabrication imperfections can drastically alter performance from ideal metrics. We explicitly show in \textbf{Section \ref{subsection:experimentallosses}} that our experimental devices have significantly lower $Q_{0}$ than the simulated value presented above while the measured $Q_{c}$ is close to the calculation detailed here. In general, fabrication imperfections will result in greater un-coupled scattering losses, lowering $Q_{0}$ and increasing $\Gamma^{0}$. Thus it is best to design a system with proper $Q_{c}$ to work for a range of $Q_{0}$ lower than simulated values.

\subsection{Experimental Device Fabrication}
\label{subsection:devicefab}
Side-coupled PCNs were fabricated to enable experimental verification of the simulated trends describing the effect of k-space overlap on coupling efficiency shown in \textbf{Section \ref{sec:Simulations}}. The devices were processed on a silicon-on-insulator (SOI) wafer with a $220 \mathrm{nm}$ thick silicon device layer and a $3 \mathrm{\mu m}$ thick buried oxide layer (SOITEC). Chips cleaved from these SOI wafers were coated with 300nm ZEP520A photoresist by spinning at 6000rpm for 45s. A JEOL93000FS tool was utilized to perform electron beam lithography. The photoresist was patterned with an electron beam at 100kV and 400 $\mathrm{\mu C/cm^{2}}$ areal dosage. Patterns were developed after exposure with gentle agitation in xylenes for 30s followed by a rinse in isopropyl alcohol. The photoresist pattern was then transferred to the SOI device layer via reactive ion etching using an Oxford PlasmaLab100. Reactive ion etching was carried out with C$_{4}$F$_{8}$/SF$_{6}$/Ar gases. Samples were cleaved after fabrication to expose the edges of the feeding waveguides to enable characterization in an end-fire coupling setup which is described in \textbf{Section \ref{subsection:experimentaltransmission}}. \textbf{Figure \ref{fig:ExpResults}(a)} shows a scanning electron microscope (SEM) image of one of the side-coupled PCNs studied in this work. 

To efficiently couple into PCN cavities with $k_{wvg} \sim k_{cav}$ and explore the effects of k-space overlap on coupling, we fabricated two, nominally identically, sample sets consisting of uniformly designed PCN cavities with feeding bus waveguide widths ranging from $w = 380 \mathrm{nm}$ to $w = 580 \mathrm{nm}$ in 10nm increments, and a fixed coupling gap size of $g = 400\mathrm{nm}$. The range of $w$ fabricated centered around the value of $w = 480$, where high resolution MODE Solutions simulations predicted $k_{cav} = k_{wvg}$. A 100nm range of widths greater and less than $w = 480$ were fabricated to account for fabrication deviations. The coupling gap $g$ = 400nm was chosen as a result of the preliminary simulations described in \textbf{Section \ref{subsec:Estimation}}.

\subsection{Experimental Transmission Data}
\label{subsection:experimentaltransmission}
\begin{figure}[t]
\center
\includegraphics[scale = .4,trim={0.1cm 0.2cm 0.2cm 2.5cm},clip,page=5]{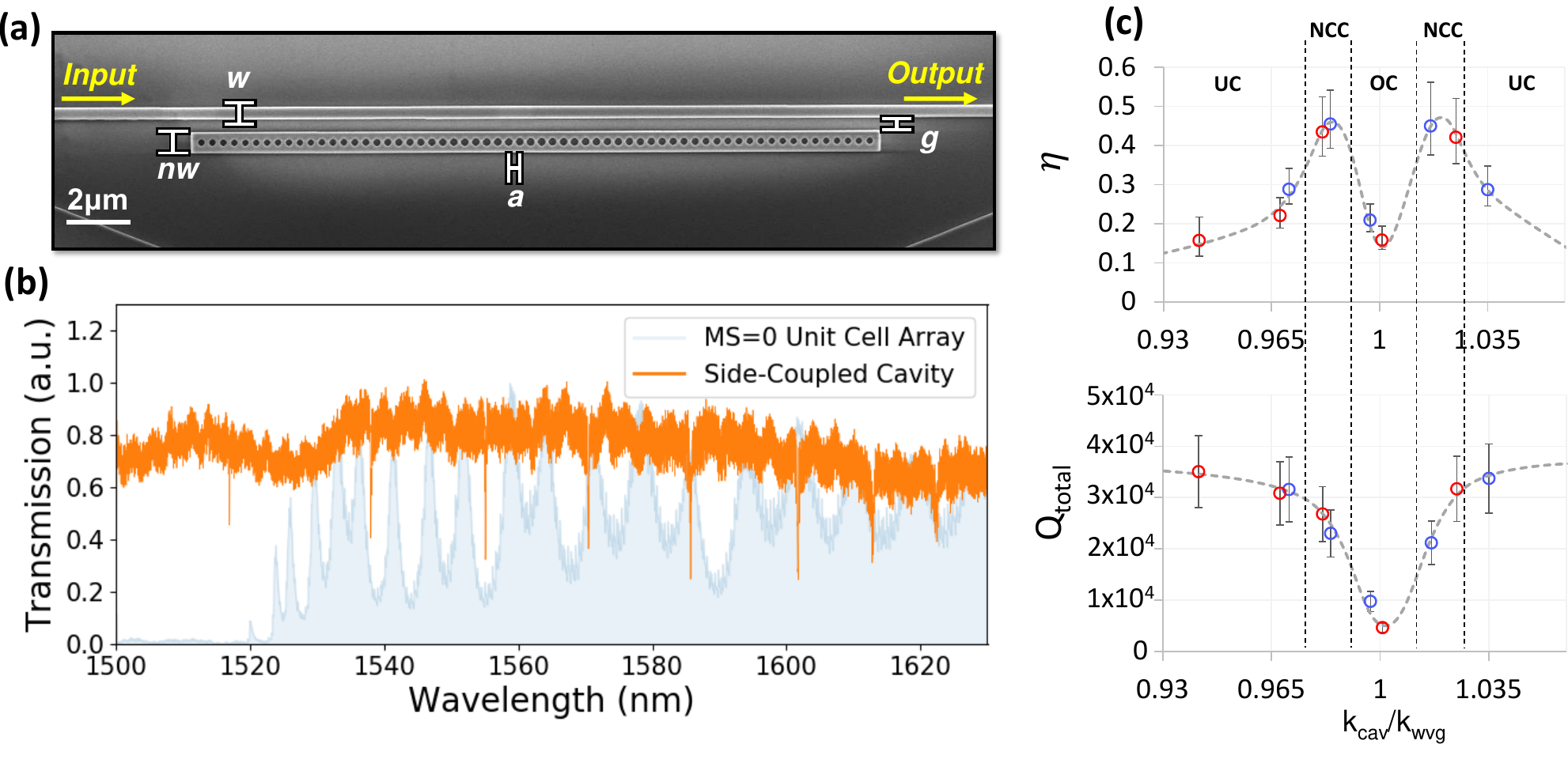}
\caption{a) SEM image of side-coupled PCN cavity fabricated for experimental testing. b) Experimental transmission spectra for a side-coupled PCN cavity and MS=0 unit cell array. The band edge of the MS=0 unit cell array lies close to the side-coupled cavity resonance at $\lambda = 1518$nm, identifying it as the fundamental PCN resonance mode. c) Experimental data for $\eta$ and $Q_{total}$ of the fundamental resonance in side-coupled PCN cavities. Blue and red data points represent data taken from two separately fabricated sample sets. The dashed grey lines are cubic spline interpolated fits to act as a guide to the eye. Under-coupled, over-coupled and near critically coupled regions are denoted by \textbf{UC}, \textbf{OC} and \textbf{NCC}, respectively.}
\label{fig:ExpResults}
\end{figure}

The transmission spectra of the side-coupled PCN cavities were measured by coupling near-infrared light from a tunable laser (1500 to 1630nm, Santec TSL-510) into and out from the bus waveguides using polarization-maintaining lensed fibers (OZ Optics Ltd.), and detecting the output light intensity with a fiber-coupled avalanche photodiode photoreceiver (Newport 2936-C). TE optical polarization was utilized for resonance excitation and to acquire measured transmission spectra. Reflections at the ends of the waveguides from the fiber-coupled measurement method are the main source of noise in the transmission spectra of \textbf{Fig. \ref{fig:ExpResults}(b)} due to Fabry-Perot interference. 

In order to verify that the fundamental resonance of the PCN cavities was measurable in the transmission spectra, a separate PCN structure was fabricated to determine the approximate location of the band edge corresponding to the central cavity unit cell of the PCN cavities. This separate PCN comprised an array of air holes possessing the same unit cell geometry as that of the central cavity unit cell of the PCN cavities under test. The band edge of this PCN was located near 1520nm as seen in \textbf{Fig. \ref{fig:ExpResults}(b)} for the spectrum labeled "MS = 0 Unit Cell Array". According to the deterministic design method, the fundamental resonance of a PCN cavity is located near the band edge wavelength corresponding to its central cavity unit cell. Experimental measurement of this band edge allows us to predict and verify the fundamental resonance position despite deviation in the fabricated PCN geometry from simulated design.

The metrics of $\eta$ and $Q_{total}$ of the side-coupled PCN cavities with different bus waveguide widths are summarized in \textbf{Fig. \ref{fig:ExpResults}(c)}.  $\eta$ was characterized by the relative transmission intensity of the resonance compared to the local baseline transmission intensity. $Q_{total}$ was extracted from transmission spectra via Lorentzian peak fitting; we note that the fundamental resonance wavelength was similar for all measured PCN cavities fabricated on the same chip, varying from 1513nm $ - $ 1521nm and 1518nm $ - $ 1530nm for the devices represented by the red and blue data points, respectively, in \textbf{Fig. \ref{fig:ExpResults}(c)}. These slight resonance wavelength variations are due to minor fabrication variations between devices, but all resonance wavelengths are near the expected band edge position. In order to highlight the importance of k-space overlap, the relative resonance depth and $Q$ of each of the measured side-coupled PCN cavities was plotted against the ratio of $k_{cav}$ to $k_{wvg}$. Following Eqns. (\ref{eqn:kcav}) and (\ref{eqn:wvgeff}), $k_{cav}$ was calculated from the PCN period, $a$, and $k_{wvg}$ was calculated from the measured resonance wavelength and the effective index estimated from Lumerical MODE Solutions simulations that considered bus waveguides with the designed widths, $w$. 

The data in \textbf{Fig. \ref{fig:ExpResults}(c)} show that tuning of the bus waveguide wave-vector via a change in $w$ enables control of the coupling regime of the side-coupled PCN cavity independent of coupling gap. As expected, large wave-vector mismatch, $\nicefrac{k_{cav}}{k_{wvg}} \gg 1$ or $\nicefrac{k_{cav}}{k_{wvg}} \ll 1$, leads to low energy exchange and under-coupling of the resonator, as characterized by a low $\eta$ and high $Q_{total}$ of the measured fundamental transmission resonance. Close wave-vector matching, $\nicefrac{k_{cav}}{k_{wvg}} \approx 1$, leads to drastically increased energy exchange; however, the lowered $Q_{total}$ and small $\eta$ are indicative of the over-coupled regime. A slight mismatch between $k_{cav}$ and $k_{wav}$ is necessary in this case to achieve near-critical coupling conditions. 

The experimental trends shown in \textbf{Fig. \ref{fig:ExpResults}(c)} are in excellent agreement with the simulated data in \textbf{Fig. \ref{fig:Concept}(b)}. The transmission spectrum of a near-critically coupled, side-coupled PCN cavity ($\nicefrac{k_{cav}}{k_{wvg}}$ = 0.98, $w$ = 410nm) with multiple high contrast resonances is shown in \textbf{Fig. \ref{fig:ExpResults}(b)}. The resonance at $\lambda = 1518$nm corresponds to the fundamental mode of the PCN cavity with measured $Q_{total} = 2.6 \times 10^{4}$ and $ \eta = 0.45 $.

\subsection{Calculation of Experimental Device Losses}
\label{subsection:experimentallosses}
\begin{figure}[b]
\center
\includegraphics[scale = .4,trim={0.4cm 4cm 2cm 2.5cm},clip,page=6]{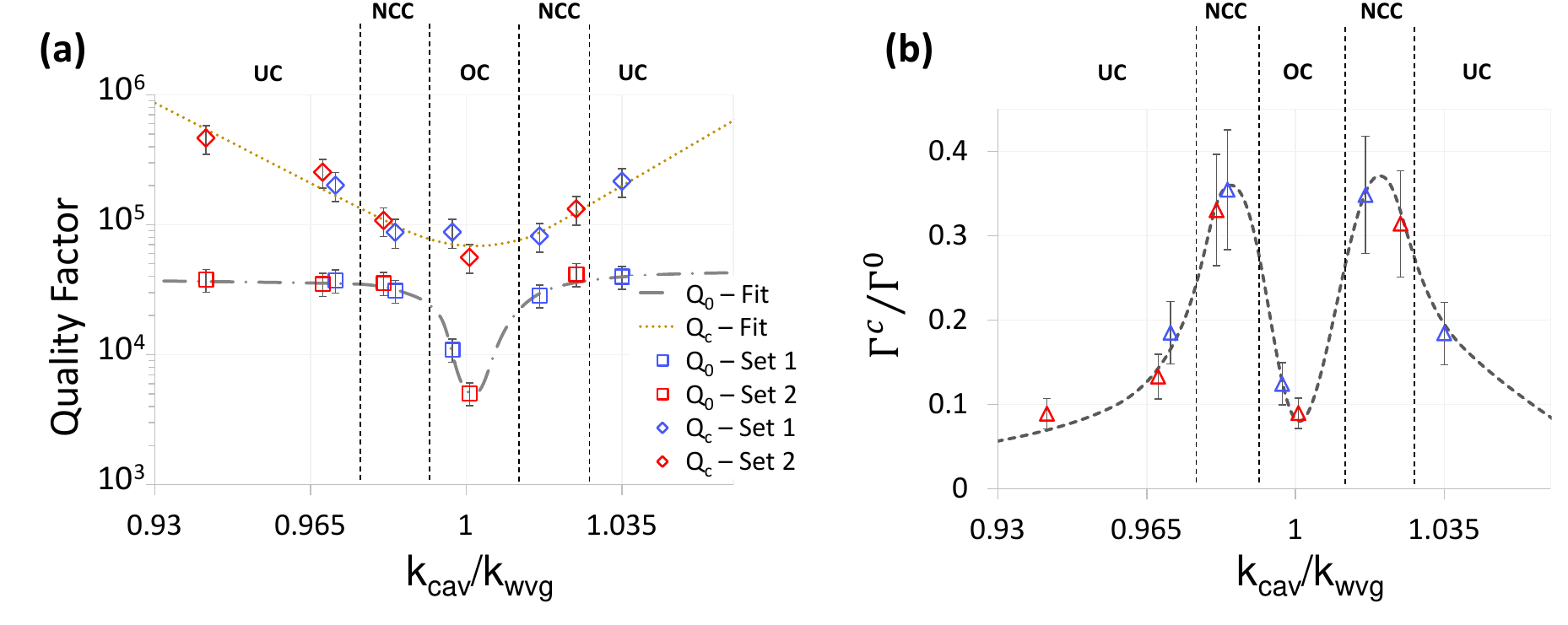}
\caption{Plots of $Q_{0}$, $Q_{c}$ and $\nicefrac{\Gamma^{c}}{\Gamma^{0}}$ calculated from experimental data. Blue and red data points represent data taken from two separately fabricated sample sets. Under-coupled, over-coupled and near critically coupled regions are denoted by \textbf{UC}, \textbf{OC} and \textbf{NCC}, respectively. a) Data for $Q_{0}$ and $Q_{c}$. A clear decrease in $Q_{0}$ can be observed near $\nicefrac{k_{cav}}{k_{wvg}} = 1$. b) Data for $\nicefrac{\Gamma^{c}}{\Gamma^{0}}$. The trend in $\nicefrac{\Gamma^{c}}{\Gamma^{0}}$ mimics the trend in $\eta$ plotted in \textbf{Fig. \ref{fig:ExpResults}(c)}}
\label{fig:QcQ0andGamma}
\end{figure}
We attribute the difference in the experimentally acquired $\eta$ for near critically coupled devices and the value acquired from simulations in \textbf{Section \ref{subsec:Estimation}} to increased intrinsic losses in fabricated devices. The higher intrinsic losses most likely stem from fabrication imperfections. We can confirm this assumption by back-calculating $Q_{0}$ and $Q_{c}$ from the experimentally measured $Q_{total}$ and $\eta$. By using the quadratic equation on Eqn. (\ref{eqn:etasidecoupled}) to solve for $\nicefrac{\Gamma^{c}}{\Gamma^{0}}$ and re-arranging Eqn. (\ref{eqn:GammaratiofromQratio}) to solve for $Q_{0}$, we get the expressions
\begin{equation}
\frac{\Gamma^{c}}{\Gamma^{0}} = \frac{-2\cdot\left(1-\eta\right)+\sqrt{4\cdot\left(1-\eta\right)^{2}+4\eta\left(1-\eta\right)}}{2\left(1-\eta\right)}
\label{eqn:GammaratiofromExperiment}
\end{equation}
\begin{equation}
Q_{0}=Q_{total}\left(\frac{\Gamma^{c}}{\Gamma^{0}}+1\right) .
\label{eqn:QintrinsicfromExperiment}
\end{equation}
Utilizing these equations and Eqn. (\ref{eqn:GammaratiofromQratio}), we can calculate $\nicefrac{\Gamma^{c}}{\Gamma^{0}}$, $Q_{0}$ and $Q_{c}$ for the devices measured in our experimental data. This data is plotted in \textbf{Fig. \ref{fig:QcQ0andGamma}}.

Fabricated devices in the under-coupled and near-critically coupled regime are calculated to have $Q_{0}\sim 3\times10^{4}$ - $4\times10^{4}$. For these configurations, $Q_{c}$ varies from $4.3\times10^{5}$ at the most under-coupled down to $8.2\times10^{4}$ in the near critically coupled region. The lowest value of $Q_{c}$ in the near critically coupled region closely matches the simulated $Q_{c} = 8.1 \times 10^{4}$ from the calculation in \textbf{Section \ref{subsec:Estimation}}. The data points in the over-coupled region show a drop in $Q_{0}$ to between $5\times10^{3}$ and $1\times10^{4}$ while the $Q_{c}$ goes down to a minimum of $5.6\times10^{4}$. This data explicitly shows over-coupling phenomena as defined in \textbf{Section \ref{subsection:computationalresults}}.

The importance of the metric $\nicefrac{\Gamma^{c}}{\Gamma^{0}}$ can also be clearly seen in \textbf{Fig. \ref{fig:QcQ0andGamma}(b)} as its trend closely matches the trend in $\eta$ from \textbf{Fig. \ref{fig:ExpResults}(c)}. By increasing the ratio of $\nicefrac{\Gamma^{c}}{\Gamma^{0}}$, we can directly improve $\eta$ for the the side-coupled PCN.

The calculation of experimental losses additionally shows that fabricated devices demonstrate lower $Q_{0}$ than simulated values while $Q_{c}$ are measured at values close to simulated results in \textbf{Section \ref{subsec:Estimation}}. We believe that fabrication imperfections which increase roughness and disorder reduce $Q_{0}$ more significantly than $Q_{c}$, making it more important to not only (1) \emph{design side-coupled PCN cavities to have high $Q_{0}$} but also (2) \emph{make $Q_{c}$ low enough to work with the range of $Q_{0}$ attainable by fabrication so a usable $\eta$ may be achieved}.

\section{Conclusion}
Side-coupled PCN cavities demonstrate attractive capabilities such as serial integration, add/drop filtering and multiplexing which are not readily attainable from in-line devices. To improve the utilization of side-coupled PCN cavities, we outline design guidelines (\textbf{Section \ref{subsec:designstrategy}}) that can enable critical coupling by appropriately tuning the ratio of coupling losses to intrinsic losses in a side-coupled PCN cavity. 

By adjusting the coupling gap and bus waveguide width we are able to practically change the overlap between the PCN cavity and bus waveguide modes in both physical and k-space. Changing this overlap enables tuning of both the coupling and intrinsic losses of the resonator with a significant degree of independence, enabling high $\eta$. While most work assumes that maximal k-space overlap, or phase matching, of devices should enable optimal device coupling efficiency, we show for our particular devices that phase matching leads to over-coupling phenomena, as k-space matching reduces $Q_{0}$ compared to $Q_{c}$, lowering $\nicefrac{\Gamma^{c}}{\Gamma^{0}}$. Therefore, a slight offset in k-space between the cavity and bus waveguide can yield the maximum coupling efficiency. By introducing a small offset in k-space, we realize a critically coupled PCN with calculated $\eta \sim 0.995$ and $Q_{total} = 7.5 \times 10^{4}$ in simulation and a near-critically coupled PCN with $\eta = 0.45$ and $Q_{total} = 2.6 \times 10^{4}$ in experiment. The generality of the analysis presented in this work should enable its application toward highly efficient coupling between other types of advanced photonic devices.

\section*{Funding}
 National Science Foundation (ECCS1407777); National Science Foundation Graduate Research Fellowship (F. O. Afzal).

\section*{Acknowledgments}
Fabrication of the photonic crystal nanobeam structures was conducted at the Center for Nanophase Materials Sciences, which is a DOE Office of Science User Facility. Scanning electron microscopy imaging was carried out in the Vanderbilt Institute of Nanoscale Science and Engineering. The authors thank Dr. Shuren Hu for helpful discussions. The authors also gratefully acknowledge Dayrl Briggs, Kevin Lester, Dr. Ivan Kravchenko, Dr. Scott Retterer and Dr. Kevin Miller for helpful discussions and assistance in fabrication. The authors would additionally like to thank Dr. Fabrice Raineri and Dr. Alex Bazin for their insightful discussions on resonator coupling, which has greatly enriched the discussion in this paper.

\section*{Disclosures}
The authors declare that there are no conflicts of interest related to this article.

%%%%%%%%%%%%%%%%%%%%%%% References %%%%%%%%%%%%%%%%%%%%%%%%%

\bibliographystyle{osajnl}
\bibliography{Bibliography}

\end{document}